\documentclass{ifacconf}
\usepackage{graphicx}      
\usepackage{natbib}
\usepackage{amsmath}
\usepackage{amsfonts}
\usepackage{amssymb}
\def\dis{\displaystyle}
\begin{document}
\begin{frontmatter}
\title{ High Fidelity Quantum State Transfer by Pontryagin Maximum Principle \thanksref{footnoteinfo}}

\thanks[footnoteinfo]{The authors acknowledge the support of FCT for the grant 2021.07608.BD, the ARISE Associated Laboratory LA/P/0112/2020, and the R$\&$D Unit SYSTEC - Base - UIDB/00147/2020 and Programmatic - UIDP/00147/2020 funds, and also the support of projects SNAP - NORTE-01-0145-FEDER-000085, HARMONY - NORTE-01-0145-FEDER-031411, funded by ERDF | COMPETE2020 | FCT/MEC | PT2020 | NORTE2020, POCI and PIDDAC.}
\author[first]{Nahid Binandeh Dehaghani}
\author[first]{Fernando Lobo Pereira}
\address[first]{SYSTEC - Research Center for Systems and Technologies, FEUP - Faculty of Engineering, Porto University, Rua Dr. Roberto Frias sn, i219, 4200-465 Porto, Portugal (e-mail: nahid@ fe.up.pt, flp@fe.up.pt).}

\begin{abstract}                
High fidelity quantum state transfer is an essential part of quantum information processing. In this regard, we address the problem of maximizing the fidelity in a quantum state transformation process satisfying the Liouville-von Neumann equation. By introducing fidelity as the performance index, we aim at maximizing the similarity of the final state density operator with the one of the desired target state. Optimality conditions in the form of a Maximum Principle of Pontryagin are given for the matrix-valued dynamic control systems propagating the probability density function. These provide a complete set of relations enabling the computation of the optimal control strategy.
\end{abstract}

\begin{keyword}
quantum optimal control, quantum fidelity measures, Pontryagin maximum principle
\end{keyword}
\end{frontmatter}
\section{Introduction}
Quantum control theory, \citep{REVIEW}, studies how the dynamics of an atomic or subatomic-level system can be manipulated by means of appropriate external electromagnetic fields or forces, designated by control, in order to maximize a given performance criterion to the system. For many quantum control protocols, the control law is required to be open-loop, that is, only a function of time and does not require state feedback data provided, for example, by measurements. In this context, optimal control theory provides powerful tools, \citep{glaser2015training}. By means of Quantum Optimal Control (QOC) theory, it is possible to formulate quantum control problems in order to seek a set of admissible controls satisfying the system dynamics while minimizing a cost functional in order to obtain a control law. In other words, QOC aims to compute the shape and sequence of control pulses to achieve a given task in an optimum way. Moreover, the high versatility of the optimal control problem formulation provides advantages such as the incorporation of diverse experimental constraints or limitations, while the optimality leads to the exploitation of physical limits of the driven dynamics, \citep{boscain2021introduction}. The investigation of optimal controls to quantum systems has been done through adaptation of traditional optimal control tools such as the variational method, \citep{kogut2011optimal}, Pontryagin maximum principle, \citep{lin2019application, yang2017optimizing, d2001optimal}, convergent iterative algorithms, \citep{wilhelm2020introduction}, among others.

The power of QOC has been instrumental in the design of several experiments, such as preparation of motional states of Bose-Einstein condensate with optimized control sequences, \citep{van2014interferometry}, and loading improvement of ultracold atoms in an optical lattice, \citep{rosi2013fast}, to name just a few. Pulses computed by using QOC allowed for error-resistant single qubit gates, \citep{timoney2008error}. QOC of a single qubit also led to the design of high-dynamic-range imaging of nanoscale magnetic fields, \citep{haberle2013high}. Virtual transitions and leakage outside the qubit manifold was also overcome in quantum processors based on superconducting circuits by means of QOC results, \citep{lucero2010reduced, motzoi2009simple}. QOC has proved to be a key tool for quantum engineering in complex Hilbert spaces, \citep{larrouy2020fast}. These experiments and several others were preceded by a large number of theoretical considerations on how QOC may impact on the control design of quantum operations including state preparation, \citep{rojan2014arbitrary, gunther2021quantum}, ranging from squeezed states, \citep{grond2009optimizing}, and cluster-states, \citep{fisher2009time}, to many-body entangled state, \citep{caneva2012entanglement}, state transfer problem, \citep{ying2016optimal, zhang2016optimal, guo2018optimal}, and quantum gate synthesis, \citep{huang2014optimal, chou2015optimal, berrios2012high}.

Amongst the above-mentioned problems, quantum state transfer has received considerable attention due to the high importance of trustworthy information transfer in quantum networks. In this regard, high-quality quantum information transport is crucial for practical models used in quantum computation, \citep{divincenzo1995quantum}. A protocol for performing fast and high fidelity quantum state transfer in quantum spin $\dis -\frac{1}{2}$ chains has been proposed in \citep{zhang2016optimal}, by using QOC based on the Krotov algorithm. Another research has shown that QOC allows a success rate of more than $98\%$ in an arbitrary state transfer process for a non-Markovian system implemented with an optimal control, \citep{ying2016optimal}. However, despite a number of researches based on QOC algorithms, the Pontryagin Maximum Principle is still far from being fully exploited for quantum systems, and, even more for the case when the dynamics described the evolution of the matrix-valued probability density function are considered. In this context, we consider a closed system described by Liouville-von Neumann equation, and formulate the control problem of quantum state transfer aiming at the maximization of the fidelity criterion. Moreover, we derive a shooting algorithm to solve the two point boundary value problem entailed by the Maximum Principle of Pontryagin by using a multiple shooting method.

The paper is organized as follows. In Section 2, after a brief description of pure and mixed quantum states, we describe the models used for closed quantum systems. The current state of the art concerning the use of the Pontryagin Maximum Principle in the context of quantum systems are described in Section 3. In the next section, we formulate a quantum state transfer in the context of optimal control subjected to Liouville-von Neumann equation with the aim of maximizing fidelity. In section 5, we present an iterative algorithm to solve the optimal control problem. The article ends with brief conclusions on the reported research as well as a brief overview on prospective research challenges.

\section{Quantum Control Modelling}

In the mathematical formulation of quantum mechanics, the state of a closed quantum system is described by a unit vector $\left| \psi \right\rangle$ in a complex Hilbert space $\mathbb{H}$ by using the Dirac representation, \citep{dirac1981principles}. In quantum information theory, information is coded by means of qubits, which can be is represented by
\begin{equation}\label{1}
\left| \psi  \right\rangle =\cos \frac{\theta }{2}\left| 0 \right\rangle +{{e}^{i\varphi }}\sin \frac{\theta }{2}\left| 1 \right\rangle
\end{equation}
where $\theta \in \left[ 0,\pi  \right]$, and $\varphi \in \left[ 0,2\pi  \right]$. Here, $\left| 0 \right\rangle$, and $\left| 1 \right\rangle$ correspond to the states $0 $ and $1$ for a classical bit, respectively, \citep{paul2007quantum}. Such quantum state represented by the wave function $\left| \psi \right\rangle$ is called a pure state.

In the density operator formalism, quantum states are described by density operators $\rho: \mathbb{H} \to \mathbb{H}$ on the system’s Hilbert space $\mathbb{H}$. For an ensemble $\left\{ {{p}_{j}},\left| {{\psi }_{j}} \right\rangle  \right\}$ of pure states, the density operator is defined as
\begin{equation}\label{2}
 \rho \equiv \sum\limits_{j}{{{p}_{j}}\left| {{\psi }_{j}} \right\rangle \left\langle  {{\psi }_{j}} \right|}
\end{equation}
in which $\left\langle  {{\psi }_{j}} \right|={{\left( \left| {{\psi }_{j}} \right\rangle  \right)}^{\dagger }}$ and $\sum\limits_{j}{{{p}_{j}}=1}$, \citep{cong2014control}.
For a pure quantum state, $\rho =\left| \psi  \right\rangle \left\langle  \psi  \right|$ and $tr\left( {{\rho }^{2}} \right)=1$, while for a mixed state $tr\left( {{\rho }^{2}} \right)<1$. Here $tr\, A $ denotes the trace of the matrix $A$. An arbitrary state $\rho$ of a qubit can be written
as a linear combination of the so-called Pauli matrices $\vec{\sigma}=(\sigma_x,\sigma_y,\sigma_z)$, which provide a basis for $2\times 2$ self-adjoint matrices as
\begin{equation}\label{3}
\rho =\frac{1}{2}\left( I+\vec{r}.\vec{\sigma}\right)
\end{equation}
where the real vector $\textbf{r}=\left( {{r}_{x}},{{r}_{y}},{{r}_{z}} \right)$ forms the coordinates of a point within the Bloch Sphere, and
$I$ indicates the $2\times2$ identity matrix. Hence, for pure states $\left\| r \right\|=1$, while $\left\| r \right\|<1$ represents mixed states, \citep{wilde2011classical}.

There are several approaches for modelling a quantum system to be controlled. In the Schrödinger model of quantum mechanics, bilinear models, including Schrödinger and quantum Liouville equations, are used to describe closed quantum systems. Schrödinger equation implies the evolution of the state vector $\left| \psi(t)  \right\rangle$ as
\begin{equation}\label{4}
	i\hbar \frac{\partial }{\partial t}\left| \psi (t) \right\rangle ={H(t)}\left| \psi (t) \right\rangle, \quad \left| \psi (t) \right\rangle_{|t=0} =\left| {{\psi }_{0}} \right\rangle
\end{equation}
where ${H(t)}$, the Hamiltonian of the system, is a Hermitian operator on $\mathbb{H}$, and $\hbar$ is the reduced Planck’s constant, considered as a unit for convenience. The system control can be realized by a set of control functions $u_k(t)\in \mathbb{R}$, which is coupled to the quantum system via time independent interaction Hamiltonians ${{H}_{k}}(k=1,2,\dots)$. Therefore, the total Hamiltonian $H\left( t \right)={{H}_{d}}+\sum\limits_{k}{{{u}_{k}}(t)}{{H}_{k}}$ determines the controlled evolution. Here, ${{H}_{d}}$ indicates the drift Hamiltonian, and the ${{H}_{k}}$'s are the interaction Hamiltonians.

For systems in mixed states, quantum statistical ensemble can be characterized via the statistical operator $\rho \left( t \right)=U\left( t,t_0 \right)\rho \left( {{t}_{0}} \right){{U}^{\dagger }}\left( t,t_0 \right)$, in which $\rho \left( {{t}_{0}} \right)=\sum\limits_{\alpha }{{{w}_{\alpha }}\left| {{\psi }_{\alpha }}\left( {{t}_{0}} \right) \right\rangle }\left\langle  {{\psi }_{\alpha }}\left( {{t}_{0}} \right) \right|$ with ${{w}_{\alpha }}$ indicating positive weights and $\left| {{\psi }_{\alpha }}\left( {{t}} \right) \right\rangle $ being the normalized state vector evolving in time according to \eqref{4}. By differentiating  $\rho \left( t \right)$  with respect to time, the equation of motion for density operator, referred to as Liouville-von Neumann equation,
\begin{equation}\label{5}
    \dot{\rho }\left( t \right)=\frac{-i}{\hbar}\left[ {H}\left( t \right),\rho \left( t \right) \right]
\end{equation}
is obtained, where $\rho \left( t \right)$ is the variable to be controlled. Equation \eqref{5} can be written in a form analogous to the classical Liouville equation as $\dot{\rho }\left( t \right)=\mathcal{L}\left( t \right)\rho \left( t \right)$, in which $\mathcal{L}$ is Liouville super-operator, \citep{cong2014control,breuer2002theory}.

\section{Short Overview on Quantum Optimal Control}
The question that we address is: how we can drive \eqref{4} from an initial state $\left| {{\psi }_{0}} \right\rangle$ to a desired target state $\left| {{\psi }_{f}} \right\rangle$ while minimizing the cost functional? There are several methods to design the optimal controller in quantum systems, which vary according to the choice of the cost function, the construction of the Pontryagin-Hamilton function, and the computation scheme using the Maximum Principle conditions, \citep{cong2014control}.

A general optimal control problem can be formulated as follows:
Given a set $\mathcal{X}$ of state functions $x:\mathbb{R}\to {{\mathbb{R}}^{n}}$, and a set $\mathcal{U}$ of control functions $u:\mathbb{R}\to {{\mathbb{R}}^{m}}$, find the functions $x\in \mathcal{X}$ and $u\in \mathcal{U}$, which minimize a cost functional $J:\mathcal{X}\times \mathcal{U}\to \mathbb{R}$ and satisfy the dynamical constraint $\dot{x}=f(x,u)$. We can formulate any optimal control problem as a specific case of the mentioned general problem, \citep{d2021introduction}. Hence, for a system with state vector $x$ driven by controls $u$ over a certain (fixed or variable) time interval $\left[ 0,T \right]$, the scalar real-valued objective functional $J$ in the form of a problem of Bolza, is expressed by
\begin{equation}\label{6}
	J :=\varPhi \left( x\left( T \right),T \right)+\int\limits_{0}^{T}{L\left( x\left( t \right),u(t),t \right)dt}
\end{equation} with $\varPhi$ and $L$ smooth functions ${{\mathbb{R}}^{n}}\times \mathbb{R}\to \mathbb{R}, {{\mathbb{R}}^{n}}\times {{\mathbb{R}}^{m}}\times \mathbb{R}\to \mathbb{R}$, respectively. The task is to maximize $J$ subjected to the condition that the system dynamics are satisfied, with $x\left( 0 \right)={{x}_{0}}$, and $u(t)$ restricted to the set of admissible controls $\mathcal{U}$, \citep{REVIEW}.

For many practical control problems in quantum setting, optimal control theory is a powerful tool to achieve quantum control objectives, \citep{peirce1988optimal, werschnik2007quantum}. The key concept of optimal control is that the control law will be obtained via minimizing a cost function that drives system \eqref{4} from $\left| {{\psi } (0)} \right\rangle=\left| {{\psi }_{0}} \right\rangle$ to the desired $\left| {{\psi } (T)} \right\rangle=\left| {{\psi }_{f}} \right\rangle$. Let consider the problem of determining the control fields $u\in {{L}^{2}}\left( \mathbb{C},\left[ 0,T \right] \right)$ while satisfying equation \eqref{4}. Suppose that we want to fulfill the following optimal criteria:
\begin{itemize}
	\item The control sequence brings the system at time $T$ to the desired state ${{\psi }_{d}}\in {{\mathbb{C}}^{n}}$.
	\item The limited laser resources are taken into account through a minimization of the control field effort.
	\item The population of intermediate states that suffer strong environment losses may need to be suppressed.
\end{itemize}
The above mentioned constraints can be summarized in the cost functional, \citep{peirce1988optimal, james2021optimal, borzi2002optimal},
\begin{equation}\label{7}
	\begin{aligned}
		&{{J}(\left| \psi \right\rangle,u)}=\frac{1}{2}\left\| \left| \psi \left( T \right) \right\rangle -\left| {{\psi }_{f}} \right\rangle  \right\|_{{{\mathbb{C}}^{n}}}^{2}\\
		&\hspace{1,3cm}+\frac{\gamma}{2}\left\| u \right\|_{{{L}^{2}}\left( \mathbb{C},\left[ 0,T \right] \right)}^{2}
		+\frac{1}{2}\sum\limits_{j=1}^{n}{{{\alpha }_{j}}}\left\| {{\psi }_{j}} \right\|_{{{L}^{2}}\left( \mathbb{C},\left[ 0,T \right] \right)}^{2}
	\end{aligned}
\end{equation}
in which $\gamma>0$ and ${{\alpha }_{j}}\ge 0$ are the weighting factors. We also may add further constraints in the same way. The problem of optimal control for minimizing cost functional \eqref{6} subject to equation\eqref{4} under normal conditions admits a solution $\left( \bar{\psi },\bar{u} \right)\in {{H}^{1}}\left( {{\mathbb{C}}^{n}},\left[ 0,T \right] \right)\times {{L}^{2}}\left( \mathbb{C},\left[ 0,T \right] \right)$, \citep{peirce1988optimal, james2021optimal, borzi2002optimal, fattorini1999infinite}. In order to calculate the necessary optimality conditions of first order, the method of Lagrange multipliers is used. Hence, the Lagrangian function is defined as
\begin{equation}\label{8}
	L\left( \left| \psi  \right\rangle ,\left| p \right\rangle ,u \right)=J\left( \left| \psi  \right\rangle ,u \right)+\operatorname{Re}\big\langle p,i\dot{\psi }-H\psi  \big\rangle
\end{equation}
where $\left\langle \phi ,\psi  \right\rangle =\int\limits_{0}^{T}{\phi \cdot}{{\psi }^{*}}dt$. Here, “*” means the complex conjugate and “$\cdot$” is the usual vector-scalar product in ${{\mathbb{C}}^{n}}$. Consider the minimization problem
\begin{equation}\label{9}
\begin{aligned}
&L( | {\tilde{\psi }} \rangle ,\left| \tilde p \right\rangle ,\tilde u)=\underset{\left| \psi  \right\rangle \in X^{0},\left| p \right\rangle,u\in {U}}{\mathop{\inf }}\,L\left( \left| \psi  \right\rangle ,\left| p \right\rangle ,u \right)\\
&U={{L}^{2}}\left( {{\mathbb{C}}^{n}},\left[ 0,T \right] \right)\\
&X^{0}={{X}}\bigcap \left\{ \psi :\psi (0)={{\psi }_{0}} \right\}
\end{aligned}
\end{equation}
The necessary conditions for problem \eqref{9} are obtained by equating to zero the Fréchet derivative of $L$ with respect to  $\left| \psi  \right\rangle$, $\left| p \right\rangle$, and $u$, \cite{REVIEW}.

Therefore, the optimality system entails:
\begin{eqnarray}
| {\dot{\psi }}\rangle& =&-i\left( {{H}_{d}}+{{H}_{k}}u \right)\left| \psi  \right\rangle ,\;\;\left| \psi \left( 0 \right) \right\rangle =\left| {{\psi }_{0}} \right\rangle \nonumber \vspace{.2cm}\\
\left| {\dot{p}} \right\rangle &=&-i\left( H_{d}+{{H}_{k}}u \right)\left|p\right\rangle -q,\vspace{.2cm}\nonumber\\
\left| p\left( T \right) \right\rangle& =&-i\left( \left| \psi (T) \right\rangle -\left| {{\psi }_{d}} \right\rangle  \right)\label{10}\vspace{.2cm}\\
u&=&\frac{1}{\gamma}\operatorname{Re}\left[ \left| p  \right\rangle {{\left( \frac{\partial \left( {{H}_{d}}+{{H}_{k}}u \right)}{\partial {{u}_{Re}}}\left| \psi  \right\rangle  \right)}^{*}} \right]\vspace{.2cm}\nonumber\\
&&\hspace{1cm}+i\frac{1}{\gamma}\operatorname{Re}\left[ \left| p  \right\rangle {{\left( \frac{\partial \left( {{H}_{d}}+{{H}_{k}}u \right)}{\partial {{u}_{Im}}}\left| \psi  \right\rangle  \right)}^{*}} \right]\nonumber
\end{eqnarray}
where ${{q}_{j}}={{\alpha }_{j}}{{\psi }_{j}}$, and $u={{u}_{Re}}+i{{u}_{Im}}$, \citep{cong2014control, borzi2002optimal, zhu1998rapid}.

\section{Formulation of the Optimal Control Problem}
Consider the following optimal control problem $(P_1)$
\begin{eqnarray*}
    \mbox{Minimize } & &-\mathcal{F}(\rho ,\sigma ) \\
    \mbox{subject to } &&\dot{\rho}(t)=F\left( \rho(t),u(t) \right) \\
     && \rho\left( 0 \right)={{\rho}_{0}}\in {{\mathbb{R}}^{n\times n}}\\
 && u\left( t \right)\in \mathcal{U}:=\left\{ u\in {{L}_{\infty }}:u\left( t \right)\in \Omega \subset {{\mathbb{R}}^{m}} \right\}
\end{eqnarray*}

where $\dot{\rho }:=\displaystyle\frac{d\rho }{dt}$, $t\in \left[ 0,1 \right]$ determines the time variable, $\rho$ is the state variable in ${{\mathbb{R}}^{n}}\times{{\mathbb{R}}^{n}}$ supposed to satisfy the differential constraints $\dot{\rho}=F\left( \rho,u \right)$ according to Liouville-von Neumann equation \eqref{5}, and $\rho_0$ is the so-called initial quantum state. $u(\cdot)$ is the measurable bounded function termed as control. Here, we aim to maximize fidelity $\mathcal{F}$ so that the density operator $\rho$ has the maximum overlap with the target $\sigma$. The most widely-used generalization of fidelity that has been indicated in the literature is the Uhlmann-Jozsa fidelity, \citep{jozsa1994fidelity}, which represents the maximal transition probability between the purification of a pair of density matrices, $\rho$ and $\sigma$, \citep{liang2019quantum}, and is defined as
\begin{equation}\label{11}
    \mathcal{F}\left( \rho ,\sigma  \right):=\underset{\left| \psi  \right\rangle ,\left| \varphi  \right\rangle }{\mathop{\max }}\,{{\left| \left\langle \psi \left| \varphi  \right. \right\rangle  \right|}^{2}}={{\left( tr\sqrt{\sqrt{\rho }\sigma \sqrt{\rho }} \right)}^{2}}
\end{equation}
satisfying the following properties, \citep{liang2019quantum},
\begin{itemize}
\item[i] $\mathcal{F}(\rho ,\sigma )=\mathcal{F}(\sigma ,\rho )$\vspace{.1cm}
\item[ii] $0\le \mathcal{F}\left( \rho ,\sigma  \right)\le 1 \quad \forall \;\rho ,\,\sigma, \quad \mathcal{F}\left( \rho ,\rho  \right)=1$\vspace{.1cm}
\item[iii] $\mathcal{F}\left( \rho ,\sigma  \right)=tr\left( \rho \sigma  \right)$ if either $\rho$ or $\sigma$ is a pure state.\vspace{.1cm}
\item[iv] $F\left( U\rho {{U}^{\dagger }},U\sigma {{U}^{\dagger }} \right)=F\left( \rho ,\sigma  \right)$ for all unitary operations $U$.

\end{itemize}
The fidelity criteria signifies a security level for quantum state transformation or the effectiveness of quantum gate synthesis, so is of high importance in quantum systems.

\section{Necessary Conditions of Optimality in the form of a Maximum Principle}

The Pontryagin-Hamilton function ${\cal H}$ is defined for almost all $t\in [0,T]$ by introducing the adjoint variable $\pi$, which is a time-varying multiplier vector designated by costate or adjoint variable of the system. Thus,
\begin{equation}\label{13}
    {\cal H}\left( \rho ,u,\pi \right)=tr\left( \pi^\dagger F\left( \rho ,u \right) \right)
\end{equation}
According to the Pontryagin's Maximum Principle, for the optimal state trajectory $\rho^*$ and the corresponding adjoint variable, the matrix $\pi$, the optimal control $ u^*(t)$, maximizes the Pontryagin-Hamiltonian function $ {\cal H}$ , i.e., for almost all $t\in[0,T]$,
\begin{equation}\label{14}
 {\cal H}\left( \rho^*(t), u, \pi(t) \right)\leq {\cal H}\left( \rho^*(t), u^*(t),\pi (t) \right)
\end{equation}
for all admissible control values $u \in \Omega$. Additionally, the adjoint equation, and its terminal conditions imply that, Lebesgue a.e.,
\begin{equation}\label{15}
 ( -\dot\pi^\dagger(t),\dot\rho^*(t))=\nabla_{( \rho ,\pi)}  {\cal H}( \rho^*(t) ,u^*(t), \pi(t))
 \end{equation}
 \begin{equation}\label{16}
 \pi^\dagger( 1)=\nabla_\rho\mathcal{F}( \rho^*(1) ,\sigma ( 1 ))
\end{equation}
From \eqref{15} we can obtain
\begin{eqnarray}
  -\dot\pi^\dagger ( t )& =&\nabla_\rho tr( \pi^\dagger F( \rho^*(t) ,u^*(t)) )\nonumber \\
 & =&\frac{-i}{\hbar}\frac{\partial }{\partial \rho }tr( \pi^\dagger(t)\left[  H^*( t ) ,\rho( t) \right])\nonumber\\
 & =&\frac{-i}{\hbar}(\pi^\dagger (t)H^*( t)-H^*( t)\pi^\dagger(t))\nonumber\\
 &=& \frac{i}{\hbar}\left[ H^*( t ),\pi^\dagger(t) \right].\label{17}
 \end{eqnarray}
 Remark that the dependence of $H$ on $t$ is through the control variable, and, hence, $H^*(t)$ denotes $H$ evaluated at each time along the optimal control $u^*$.
It is straighforward to conclude that the differential equation \eqref{17} has the formal solution
\begin{equation}\label{18}
    \pi^\dagger( t)=e^{i\int_t^1H^*(s)ds}\pi^\dagger(1)e^{-\int_t^1H^*(s)ds}
\end{equation}
and the boundary condition at the final time is obtained by computing
\begin{equation}\label{19}
\pi^\dagger( 1)=\nabla_\rho\left( tr\sqrt{\sqrt{\rho(1) }\sigma(1) \sqrt{\rho (1) }}\right)^2
\end{equation}
for which by means of Taylor expansion at the point $\rho = I$, we have
\begin{equation}\label{20}
\sqrt{\rho }=\sum_{k=0}^\infty\frac{1}{k!}\frac{d^k}{d\rho^k}\sqrt{\rho}_{|\rho = I}( \rho -I)^k
\end{equation}
which, in turn, under some conditions, the application of the Cayley–Hamilton theorem yields, for a certain coefficients $\alpha_k$, $k=0,\ldots, n-1$, being $n$ the dimension of the square matrix $\rho$,
$$ \sqrt{\rho }=\sum_{k=0}^{n-1}\alpha_k( \rho -I)^k.$$ Thus, by using matricial calculus, we obtain
\begin{eqnarray}
\pi^{\dagger}( 1)&=&2 tr\sqrt{\rho(1)\sigma(1)}\nabla_\rho( tr\sqrt{\rho(1) \sigma(1) })\nonumber\\
 & \hspace{-.9cm}= &\hspace{-.5cm}2 tr\sqrt{\rho(1) \sigma(1) }\sum\limits_{k=0}^{n-1}{\alpha }_{k}\sum_{i=0}^{k-1}\bar \rho(1)^i\sqrt{\sigma ( 1 )}\bar \rho(1)^{k-i-1}\label{21}
\end{eqnarray}
where $ \bar \rho(1)= \rho(1)-I$.
\section{Application of the Pontryagin Maximum Principle}
Let us consider a spin $\dis -\frac{1}{2}$ particle in an invariable magnetic field $B_0$ along z-axis as the controlled system, the control magnetic fields on the x–y plane is given by
\begin{equation}
\begin{aligned}
  & \gamma {{B}_{x}}=u\left( t \right)\cos \left( \omega t+\phi  \right) \\
 & \gamma {{B}_{y}}=-u\left( t \right)\sin \left( \omega t+\phi  \right) \\
\end{aligned}
\end{equation}
where $\gamma$ is the magnetic ratio of spin particle, and $u(t)$ is a real valued number indicating the Rabi frequency of the particle. Hence, The system Hamiltonian $H(t)$, composing of the free Hamiltonian $H_d$ and the control Hamiltonian $H_c$, is expressed as
\begin{equation}\label{22}
\begin{aligned}
H\left( t \right)=&\frac{-\gamma \hbar }{2}\left( {{B}_{0}}{{\sigma }_{z}}+{{B}_{x}}\left( t \right){{\sigma }_{x}}+{{B}_{y}}\left( t \right){{\sigma }_{y}} \right)\\
    =&-\frac{\hbar }{2}\left( \begin{matrix}
   {{\omega }_{0}} & u(t) {e}^{i\left(\omega t+\phi  \right) } \\
   u(t) {{e}^{-i\left( \omega t+\phi  \right) }} & -{{\omega }_{0}}  \\
\end{matrix} \right)\\
\end{aligned}
\end{equation}
in which $\gamma {{B}_{0}}=\omega _ 0$, and ${\sigma }_{x}$, ${\sigma }_{y}$, and ${\sigma }_{z}$ are the so-called Pauli matrices given by
\begin{equation}
{{\sigma }_{x}}=\left( \begin{matrix}
   0 & 1  \\
   1 & 0  \\
\end{matrix} \right),\quad {{\sigma }_{y}}=\left( \begin{matrix}
   0 & -i  \\
   i & 0  \\
\end{matrix} \right),\quad {{\sigma }_{z}}=\left( \begin{matrix}
   1 & 0  \\
   0 & -1  \\
\end{matrix} \right)
\end{equation}

Once the matrix $H(t)$ is known, the matrix $U(t)$ of this system can be computed by
\begin{equation}\label{24}
U(t)={e}^{-\frac{i}{\hbar}\int_0^tH(s)ds}
\end{equation}
The system dynamics are given by the Liouville equation whose solution is analogous to the one of the adjoint system considered in a previous section, that is,
\begin{equation}\label{25}
\rho (t)=U\left( t \right)\rho (0){{U}^{\dagger }}\left( t \right)
\end{equation}
Here, we consider a state transfer problem from $\left| 0 \right\rangle$ to $\left| 1 \right\rangle$, so $\rho \left( 0 \right)=\left( \begin{matrix}
   1 & 0  \\ 0 & 0  \\ \end{matrix} \right)$ and $\sigma \left( 1 \right)=\left( \begin{matrix} 0 & 0  \\ 0 & 1  \\ \end{matrix} \right)$.

In this section, we present a time discretized computational scheme to solve the optimal control problem associated with $(P_1)$ by using an indirect method based on the Maximum Principle.
Let $N$ be the number of discrete time subintervals. Given the smooth properties of the problems data, let us consider a uniform discretization. Thus, we consider the $N$ points $\dis t_k = \frac{k}{N}$ for $ k = 0,\ldots, N-1$ and denote the value of any function $f(t_k)$ by $f_k$. We consider the both the system dynamics, and the adjoint differential equations approximated by a first order Euler approximation. Higher order methods yield better approximations but our option is made to keep the presentation simple. Let $j=0,\ldots$ be the iterations counter, and we denote the $j^{th}$ iteration of the function $f$ at time $t_k$ by $f_k^j$. The proposed algorithm is as follows:
\begin{itemize}
\item[] \hspace{-.5cm}{\bf Step 1 - Initialization}. \vspace{.1cm}

\noindent Let $j=0$, and initialize the values of $u_k^j$ for $k=1,\ldots,N-1$, $\omega^j$, and $\phi^j$ in (\ref{22}).\vspace{.3cm}

\item[] \hspace{-.5cm}{\bf Step 2 - Computation of the state trajectory}. \vspace{.1cm}

\noindent For $k=0,\dots,N-1$, let $\dis \rho _{k+1}^j=\rho_k^j+\frac{1}{k}\dot\rho_k^j $. \vspace{.3cm}

\item[] \hspace{-.5cm}{\bf Step 3 - Computation of the adjoint trajectory}.\vspace{.1cm}

\noindent Compute $\pi_N^j $ by using \eqref{21} with $\rho _N^j$ computed in Step 2.\vspace{.1cm}

\noindent Compute $\pi_0^j $ by using the discretized version of \eqref{18}, that is, $\dis\pi_0^j = U^j{\pi_N^j}{U^j}^\dagger $ 
where $\dis  U^j = e^{i\sum_{k=0}^{N-1}\frac{1}{N}H_k^j}$, being $H_k^j $ the Hamiltonian of the system dynamics with the value control $u^j$ at time $t_k$.\vspace{.1cm}

\noindent For $k=0,\dots,N-1$, let $\dis \pi _{k+1}^j=\pi_k^j+\frac{1}{k}\dot\pi_k^j $, \vspace{.3cm}

\item[] \hspace{-.5cm}{\bf Step 4 - Computation of the Pontryagin Hamilton function}\vspace{.1cm}

\noindent For $k=0,\dots,N-1$, let $$\dis \mathcal{H}_k^j(u,\omega,\phi)= tr(-i {\pi^j_k}^\dagger[H(u,\omega,\phi,t_k),\rho_k^j])$$  
where $\dis H(u,\omega,\phi,t_k)$, from \eqref{22}, is given by $$\dis -\frac{1 }{2}\left( \begin{matrix}
   {{\omega }_{0}} & u {e}^{i\left(\omega t_k+\phi  \right) } \\
   u {{e}^{-i\left( \omega t_k+\phi  \right) }} & -{{\omega }_{0}}  \\
\end{matrix} \right)$$

\vspace{.2cm} 
\item[] \hspace{-.5cm}{\bf Step 5: Update the control function.}\vspace{.1cm}

\noindent For $k=0,\ldots, N-1$, compute the values $u_k^{j+1}$, $\omega^{j+1}$, and $\phi^{j+1}$ that maximize the map \vspace{-.1cm}
$$ (u,\omega,\phi) \to \mathcal{H}_k^j(u,\omega,\phi).$$

\vspace{.2cm} 
\item[] \hspace{-.5cm}{\bf Step 6: Stopping test}.\vspace{.1cm}

\noindent For given small positive numbers $\varepsilon_\omega$, $\varepsilon_\phi $, and $\varepsilon_u$ (tolerance errors), check whether all the following inequalities hold:
\begin{eqnarray*}
&&\hspace{1.5cm} | \omega^j-\omega^{j+1}|<\varepsilon_\omega \\
&& \hspace{1.5cm}| \phi^j-\phi^{j+1}|<\varepsilon_\phi \\
&& \hspace{.5cm}  \max_{k=0,\ldots,N-1}\{|u_k^j-u_k^{j+1} |\}<\varepsilon_u 
\end{eqnarray*}

\smallskip

\noindent If yes, let, $\omega^*= \omega^{j+1}$, $\phi^*= \phi^{j+1}$, for $k=0,\ldots,N-1$, $u^*(t_k) = u_k^{j+1}$, and exit the algorithm. 

\smallskip 

\noindent Otherwise, let $ j=j+1$, go to {\bf Step 2}.

\end{itemize}

\section{Conclusion}
In this paper we have shown how the Maximum Principle of Pontryagin can be applied in order to compute the optimal control of the problem of maximizing fidelity in the transfer of the state variable, expressed by the matrix-valued probability density function, between two given state values with the dynamics of the system the Liouville-von Neumann equation. This context is not very common neither in Optimal Control nor in Quantum Optimal Control. 
We obtained a shooting algorithm to solve the two point boundary value problem resulting from the application of the Maximum Principle of Pontryagin.
Future challenges consists in exploiting the versatility of the optimal control paradigm further by including state constraints and other types of constraints. More efficient algorithms to solve this problem will also be considered.

\bibliography{IFAC}

\end{document}